\documentclass[epj]{svjour}
\usepackage{graphics}
\begin{document}
\title{Universal behavior of baryons and mesons transverse momentum
  distributions in the framework of percolation of strings}
\author{L. Cunqueiro$^1$ 
\and J.Dias de Deus$^2$
\and E. G. Ferreiro$^1$ 
\and C. Pajares$^1$} 

\institute{$^1$Instituto Galego de F\'{\i}sica de Altas Enerx\'{\i}as and
  Departamento de F\'{\i}sica de Part\'{\i}culas, Universidade de Santiago de
  Compostela, 15782 Santiago de Compostela, Spain.\\
  $^2$ CENTRA, Instituto Superior T\'ecnico, 1049-001 Lisboa, Portugal}

\date{Agosto. 1, 2007}
\abstract{In the framework of percolation of strings, the transverse momentum
distributions in AA and hh collisions at all centralities and energies follow
a universal behavior. The width of these distributions is related to the
width of the distribution of the size of the clusters formed from the
overlapping of the produced strings. The difference between the
distributions for baryons and mesons originates in the fragmentation of
clusters of several strings which enhance the particles with higher number of
constituents. The results agree with SPS and RHIC data. The predictions for
LHC show differences for baryons compared with RHIC. At LHC energies we obtain
also a high pt suppression for pp high multiplicity events compared with pp
minimum bias. 
}

 % end of PACS codes
 %end of abstract
%
\maketitle
%%%%%%%%%%%%%%%%%%%%%%%%%%%%%%%%%%%%%%%%%%%%%%%%%%%%%%%%%%%%%%%%%%
Multiparticle production can be described in terms of color strings stretched
between the partons of the projectile and target. These strings decay into new
ones by sea $q-\overline{q}$ pair production and subsequently hadronize to produce
the observed hadrons. The color in these strings is confined to a small area
in  transverse space: $S_{1}=\pi r_{0}^{2}$ with  $r_{0}\simeq 0.2-0.3$ fm.
With increasing energy and/or atomic number of the colliding particles, the
number of exchanged strings grows and they start to overlap , forming clusters,
very much like disks in two-dimensional percolation theory. At a certain
critical density, a macroscopical cluster appears, which marks the percolation
phase transition \cite{Armesto:1996kt,Braun:2000hd}. For nuclear collisions, this density
corresponds to the value of $\eta=N_{S}\frac{S_{1}}{S_{A}}$ of
$\eta_{C}=1.18-1.5$(depending of the type of the employed profile
functions) where $N_{S}$ is the number of strings and $S_{A}$ corresponds to
the overlapping area of the nuclei. A cluster of n strings behaves as a single
string with and energy-momentum that corresponds to the sum of the
energy-momentum of the overlapping strings and with a higher color field,
corresponding to the vectorial sum of the color fields of the individual
strings. In this way, the multiplicity $<\mu_{n}>$ and the mean transverse
momentum squared $<p_{T}^{2}>_{n}$ of the particles produced by a cluster are
given by
\begin{equation}<\mu>_{n}=\sqrt{\frac{nS_{n}}{S_{1}}}<\mu>_{1} \hspace{1cm} <p_{T}^{2}>_{n}=\sqrt{\frac{nS_{1}}{S_{n}}}<p_{T}^{2}>_{1}  \label{eq:average} \end{equation}
where $<\mu>_{1}$ and $<p_{T}^{2}>_{1}$ stand for the mean multiplicity and
mean $p_{T}^{2}$ of particles produced in a single string. 
Eqs.\ref{eq:average} transform into analytical ones \cite{Braun:2000hd} in the
limit of random distributions of strings
   \begin{equation} <\mu>=N_{S} F(\eta)<\mu>_{1} \hspace{1cm}
     <p_{T}^{2}>=\frac{<p_{T}^{2}>_{1}}{F(\eta)}\label{eq:analytic} \end{equation}
where $F(\eta)=\sqrt\frac{1-e^{-\eta}}{\eta}$. If we are interested in a
determined kind of particle i, we will use $<\mu>_{1i}$, $<p_{T}^{2}>_{1i}$
$<\mu>_{ni}$ and $<p_{T}^{2}>_{ni}$ for the corresponding quantities.
The transverse momentum distributions can be written as a superposition of the
transverse momentum distributions of each cluster, $g(x,p_{T})$, weighted with
the distribution of the different tension of the clusters, i.e the distribution
of the size of the clusters, $W(x)$ \cite{Dias de Deus:2003ei,Pajares:2005kk}.
For $g(x,p_{T})$ we assume the Schwinger formula $g(x,p_{T})=exp({-p_{T}^{2} x})$
and for the weight function $W(x)$, the gamma distribution
$W(x)=\frac{\gamma}{\Gamma(k)}(\gamma x)^{k-1}exp({-\gamma x}) $
where $$\gamma=\frac{k}{<x>} \; \; \hspace{1cm} \frac{1}{k}=\frac{<x^2>-<x>^2}{<x>^2}.$$
$x$ is proportional to the inverse of the tension of each cluster, precisely
$x=1/<p_{T}^{2}>_{n}=\sqrt{\frac{S_{n}}{nS_{1}}}\frac{1}{<p_{T}^{2}>}$. $k$ is
  proportional to the inverse of the width of the distribution on $x$ and
  depends on $\eta$, the density of strings.
Therefore, the transverse momentum distribution  $f(p_{T},y)$ is
\begin{eqnarray}\frac{dN}{dp_{T}^2 dy}=f(p_{T},y)=\int_{0}^{\infty}dx W(x)
g(p_{T},x)= \nonumber \\
\frac{dN}{dy}\frac{k-1}{k}\frac{1}{<p_{T}^2>_{1i}}F(\eta)\frac{1}{(1+\frac{F(\eta)p_{T}^{2}}{k<p_{T}^2>_{1i}})^{k}}
\label{eq:spectra}\end{eqnarray}
Eq.(\ref{eq:spectra})  is valid for all densities and type of collisions. 
It only depends on
the parameters $<p_{T}>_{1i}$ and k. At low
density, there is no overlapping between strings and there are no fluctuations
on the string tension , all the clusters have one string. Therefore, k goes to
infinity, and $f(p_{T},y) \simeq exp({-\frac{p_{T}^{2}}{<p_{T}^{2}>_{1}}})$. At
very high density $\eta$, there is only one cluster formed by all the produced
strings. Again, there are no fluctuations, k tends to infinity and
$f(p_{T},y) \simeq exp({-\frac{F(\eta) p_{T}^{2}}{<p_{T}^{2}>_{1}}})$ . In
between these two limits, k has a minimum for intermediate densities
corresponding to the maximum of the cluster size fluctuations. This behavior
of k with $\eta$ \cite{Dias de Deus:2003ei,Pajares:2005kk} is related to the behavior with centrality
of the transverse momentum \cite{Adcox:2004mh} and multiplicity fluctuations \cite{Alt:2006jr}.
We observe from eq.(\ref{eq:spectra}) that
\begin{equation}\frac{dlnf(p_{T}^2,y)}{dlnp_{T}}=-\frac{2F(\eta)}{(1+\frac{2F(\eta)p_{T}^{2}}{k<p_{T}^{2}>_{i}})}
\frac{p_{T}^{2}}{<p_{T}^{2}>_{1i}} \label{eq:deriv} \end{equation} 
As $p_{T}^{2} \longrightarrow 0$, eq.(\ref{eq:deriv}) reduces to
$\frac{-2F(\eta)p_{T}^{2}}{<p_{T}^{2}>_{i}}$ while for larger $p_{T}$ it
becomes $-2k$ for all particle species. As $<p_{T}^{2}>_{1p} \ge
<p_{T}^{2}>_{1K} \ge <p_{T}^{2}>_{1\pi}$, the absolute value is larger for
pions than for kaons and for protons, in agreement with experimental data
\cite{Dias de Deus:2003ei}.

The nuclear modification factor
, defined as
$R_{AA}(p_{T})=\frac{dN^{AA}}{dp_{T}^{2}dy}/N_{coll}\frac{dN^{pp}}{dp_{T}^{2}dy}$,
reduces to the following expression at $p_{T}^{2}=0$ (we use eq.2 for $\frac{dN}{dy}$):   
 \begin{equation}
R_{AA}(0)\sim ( \frac{F(\eta')}{F(\eta)})^{2} < 1 \label{eq:limit}\end{equation}
where $\eta'$ and $\eta$ are the corresponding densities for nucleus-nucleus
collisions and pp collisions respectively. $\eta' > \eta$, thus $F(\eta') <
F(\eta)$. As $p_{T}$ increases we have
\begin{equation}
R_{AA}(p_{T})\sim
\frac{1+F(\eta)\frac{p_{T}^{2}}{<p_{T}^{2}>_{1i}}}{1+F(\eta')\frac{p_{T}^{2}}{<p_{T}^{2}>_{1i}}} (\frac{F(\eta')}{F(\eta)})^{2}
\label{eq:limit2}\end{equation}
and $R_{AA}$ increases with $p_{T}$ up to a maximum value. At larger $p_{T}$ ($\frac{p_{T}^{2}}{k<p_{T}^{2}>_{1i}}F(\eta)>1$)
\begin{equation}
R_{AA} \sim
p_{T}^{2(k(\eta)-k'(\eta'))} \label{eq:limit3}\end{equation}
At high density $k'(\eta') > k(\eta)$ and suppression of $p_{T}$ occurs.

The universal formula (\ref{eq:spectra}) must be regarded as an analytical approximation to a
process which consists on the formation of clusters of strings and their
eventual decay via the Schwinger mechanism. We do not claim to have an
alternative description valid at all $p_{T}$. It is well known that jet
quenching is the working mechanism responsible of the high $p_{T}$
suppression. This phenomena is not included in our formula which was obtained
assuming a single exponential for the decay of a cluster without a power like
tail. Our work must be considered as an interpolating way of joining smoothly
the low and intermediate $p_{T}$ region with the high $p_{T}$ region. The
suppression of high $p_{T}$ has correspondence with a modification in the
behavior at intermediate $p_{T}$ and this is what we study.
Although it describes many of the observed features of experimental
SPS and RHIC data, it is not able to explain the differences between
antibaryons(baryons) and mesons. In fact, the only different parameter between
them in formula(\ref{eq:spectra}) is the mean transverse momentum of pions and protons
produced by a single string $<p_{T}^{2}>_{1 \pi}$ and $<p_{T}^{2}>_{1 p}$
respectively. This only causes a shift on the maximum of $R_{AA}$ but keeps
the same height at the maximum contrary to the observed. However, in the
fragmentation of a cluster formed by several strings, the enhancement of the

production of antibaryons(baryons) over mesons is not only due to a mass
effect corresponding to a higher tension due to higher density of the
cluster(factor $F(\eta)$in front of $p_{T}^{2}$ in formula(\ref{eq:spectra})). In fact, the
color and flavor properties of a cluster follow from the corresponding
properties of their individual strings. A cluster composed of several
quark-antiquarks $(q-\overline{q})$ strings behaves like a ($Q-\overline{Q}$)
string, with a color Q and flavor composed of the flavor of the individual
strings. As a result, we obtain clusters with higher color and differently
flavored ends. For the fragmentation of a cluster we consider the creation
of a pair of parton complexes $Q\overline{Q}$ \cite{Amelin:1994mc}. After the
decay, the two new $Q\overline{Q}$ strings are treated in the same manner and
therefore decay into more $Q\overline{Q}$ strings until they come to objects
with masses comparable to hadron masses which are identified with observable
hadrons by combining into them the produced flavor with statistical
weights. In this way, the production of antibaryons(baryons) is enhanced with
the number of strings of the cluster. As an example, in fig.1 we show the
results for the decay of a color octet cluster and color sextet formed by
two $3-\overline{3}$ strings \cite{Amelin:1994mc}, were there is a large
enhancement of antibaryons(baryons). Notice that the enhancement of
strangeness with zero baryon number is smaller \cite{Amelin:1994mc,Armesto:2001et,Mohring:1992wm}. We
observe that the additional antiquarks (quarks) required to form an
(anti)baryon are provided by the antiquarks(quarks) of the overlapping
strings which form the cluster. In this way, the recombination and coalescence
ideas \cite{Hwa:2003ce} are naturally incorporated in our approach. In order to
take this into account in our formulas, we must modify the eqs.(\ref{eq:average}),(\ref{eq:analytic}) and (\ref{eq:spectra}). For
(anti)baryons we will consider the multiplicity per unit of rapidity to be
instead of the first equation of \ref{eq:analytic}:
\begin{equation}\mu_{\overline{B}}=N_{S}^{1+\alpha}F(\eta_{\overline{B}})\mu_{1\overline{B}}\label{eq:dndy} \end{equation}
fitting the parameter $\alpha$ to reproduce the experimental dependence of the
pt integrated  $\overline{p}$ spectra with centrality \cite{Adler:2003cb}. The
result is shown in fig.2 and the obtained values are $\alpha=0.09$ and
$\frac{\mu_{1\overline{p}}}{\mu_{1\overline{\pi}}}=\frac{1}{30}$.
$\mu_{1\overline{p}}$ and $ \mu_{1\overline{\pi}}$ are the mean multiplicity
of a single string for anitprotons and pions respectively.
It is observed that when an antibaryon is triggered, the effective number of
strings is $N_{S}^{1+\alpha}$ instead of $N_{S}$. This means that the density
$\eta$ must be replaced by $\eta_{\overline{B}}=N_{S}^{\alpha}\eta$. The
(anti)baryons probe a higher density than mesons for the same energy and type
of collision. On the other hand, from  constituent counting rules \cite{Brodsky:1973kr}
it is expected that the power-like $p_{T}$ behavior for baryons is
suppressed in one half more than mesons. 
Therefore, in (\ref{eq:spectra}) we must use for (anti)baryons $\eta_{B}$ and
the corresponding functions $F(\eta_{B})$ and $k_{B}=k(\eta_{B})+\frac{1}{2}$. Since
$\eta_{B} > \eta$ we have $F(\eta_{B})<F(\eta)$ and $k(\eta_{B})>k(\eta)$. For
peripheral collisions, $N_{S}^{\alpha}$ is smaller than for central
collisions and $\eta_{B}$ is more similar to $\eta$. Therefore, the
differences between the transverse momentum distributions are smaller as it is
shown by the experimental data. In order to compute the different $\eta$s for
different centralities we use the Monte-Carlo code of references
\cite{Amelin:1994mc,Armesto:2001et}. Their values at RHIC and LHC and their corresponding k's are
tabulated below. Note that the values of k come from the universal function
which gives the shape of the dependence of k on $\eta$ \cite{Dias de Deus:2003ei,Pajares:2005kk}. 

In our approach, we should use a different $\alpha$ for baryons and
antibaryons as far as the increase with centrality is slightly different for
both, depending also on the specific kind of baryon(antibaryon). In order to
obtain a single formula we considered strings of the same type. However, usually two
types of strings are considered. Strings $qq-q$ or $q-qq$ which stretch a diquark
of the projectile (target) with a quark of the target(projectile) and strings $q-\overline{q}$
or $\overline{q}-q$ linking quarks and antiquarks. The fragmentation of the
strings of the type $qq-q$ or $q-qq$ favors the production of baryons over
antibaryons in the fragmentation regions of the projectile and the
target. Therefore our results should be limited to the central rapidity region. On
the other hand, we have determined $\alpha$ from  the dependence of
$\frac{\overline{p}}{\pi}$ with  centrality. The large error data translate
into uncertanties in $\alpha$ of the order of $20\%$ which is of the same
order of the differences in the parameter $\alpha$ for baryons and
antibaryons. The uncertanties in $\alpha$ induce uncertanties in the
determination of $\eta_{B}$ and hence in $k_{B}$; these uncertanties are however
negligible (less than 5$\%$ even at the highest centrality).
We conclude that our single formula using the same $\alpha$ for baryons and
antibaryons is a good approximation.

On the other hand, we have assumed an additional difference between baryons
and mesons, $k_{B}=k(\eta_{B})+1/2$, from the high $p_{T}$ behavior of
eq.3. This could seem inconsistent with our approach that is limited to low
and intermediate $p_{T}$. As we have said above, our goal is not to give a full
description of data, including high $p_{T}$ but an alternative description of
intermediate $p_{T}$ suppression. In this way, this factor can be seen as a
boundary condition imposed to our approach. Notice that this factor is
independent of the centrality and therefore its influence on the ratios
$R_{CP}$ and $R_{AA}$ is not large. In fact, if we use for $k_{B}$ the values
without the factor $1/2$, $R_{CP}$ is smaller at $p_{T}=2$ GeV/c and
$p_{T}=10$ GeV/c in a  factor 1.13 and 1.25 respectively at RHIC
energies. This small reduction does not spoil the agreement with data although
a better agreement at high $p_{T}$ is obtained with the 1/2 factor.

In fig 3., we show our results for the ratio $R_{CP}$ in Au-Au
collisions defined as usual, for $(p+\overline{p})/2$ (dashed line) and neutral
pions(solid line), compared to the PHENIX \cite{Adler:2003cb} experimental
data. We find a good agreement. 
We also show the LHC prediction for pions and antiprotons. There is no change
for pions but on the contrary, the difference between antibaryons and pions is
enhanced. 
In fig.4 we show the $p_{T}$ dependence of the
ratio $\frac{\overline{p}}{\pi^{0}}$ for peripheral (dashed line) and
central (solid line) Au-Au collisions together with experimental data
\cite{Adler:2003cb}. LHC predictions are also shown. The difference only appears for
central collisions.

In fig.5 we show our results for the modified nuclear factor $R_{AA}$ for
pions (solid lines) and protons (dashed lines) for peripheral and central
collisions together with the experimental data for pions(\cite{Adler:2003qi}). Again,
the main difference arises for central collisions. 
In fig.6, this difference
between pions and protons is compared at RHIC and LHC  energies. The Cronin
effect becomes larger at LHC for protons, contrary to some
expectations\cite{Albacete:2003iq}.

The good agreement obtained with the experimental data can be understood as
two combined effects:the larger string tension of the cluster and the formation
of strong color fields\cite{Topor Pop:2007hb} and the way of fragmentation of the
clusters which enhances (anti)baryon over mesons similarly to recombination
models\cite{Hwa:2003ce}. Both effects are widely recognized as working physical
mechanisms at high densities . Both effects are naturally incorporated in the
percolation of strings approach.

The shape of $R_{AA}$ and $R_{CP}$ has nothing to do with the nucleon structure of
the nucleus an it depends essentially on the string density. One can wonder
whether in pp at LHC energies can be reached enough string densities to get a
high $p_{T}$ suppression. In fig. 7 we answer this question. It is plotted the
ratio $R_{CP}$ between the inclusive pp going to $\pi$, $k$ and $\overline{p}$
cross section for events with a multiplicity twice higher than the mean
multiplicity and the minimum bias cross section. It is observed a suppression
for $p_{T}$ larger than $3$ GeV/c. However, the reached string density in pp
collisions will not be enough to suppress the back to back jet correlations as
observed at RHIC energies for Au-Au central collisions. A straightforward
evaluation following \cite{Dias de Deus:2003ei}, gives a maximum of absorption of $\Delta
p_{T}=2$ GeV/c
for a number of strings of the order of 25-30 corresponding to events with a
multiplicity twice the minimum bias multiplicity.

We thank Ministerio de Educaci\'on y Ciencia of Spain under project
FPA2005-01963 and Conselleria de Educaci\'on da Xunta de Galicia for financial
support. We thank N.Armesto, C.Salgado and Y.Shabelski for discussions.
\begin{table}
\begin{center}
\begin{tabular}{|c|c|c|c|c|} \hline
\ \ Centrality \ \
&$\eta$ \ \

& $k$  \ \ & $\eta_{B}$ \ \ & $k_{B}$ \ \ \\ \hline
\multicolumn{5}{|c|} {RHIC} \\ \hline

 Au-Au $0-10\%$  & 2.69 & 3.97 & 5.17 & 4.82 \\ \hline

 Au-Au $60-92\%$ & 0.9 & 3.58 & 1.24& 4.16 \\ \hline

 Au-Au $80-92\%$ & 0.6 & 3.55 & 0.75 & 4.06 \\ \hline

 pp & 0.4 & 3.60 & 0.48 & 4.07 \\ \hline
 \multicolumn{5}{|c|} {LHC} \\ \hline

 Au-Au $0-10\%$ & 4.85 & 4.07 & 9.8 & 4.99 \\ \hline

 Au-Au $60-92\%$ & 1.62 & 3.56 & 2.34 & 4.21 \\ \hline

 Au-Au $80-92\%$ & 1.08 & 3.44 & 1.43 & 4.02 \\ \hline

 pp & 0.72 & 3.38 & 0.92 & 3.90 \\ \hline
\end{tabular}
\end{center}
\caption{Density of strings and the corresponding k values for mesons and
  barions at RHIC and LHC.}
\end{table}

\begin{figure}
\resizebox{0.5\textwidth}{0.3\textheight}{
\includegraphics{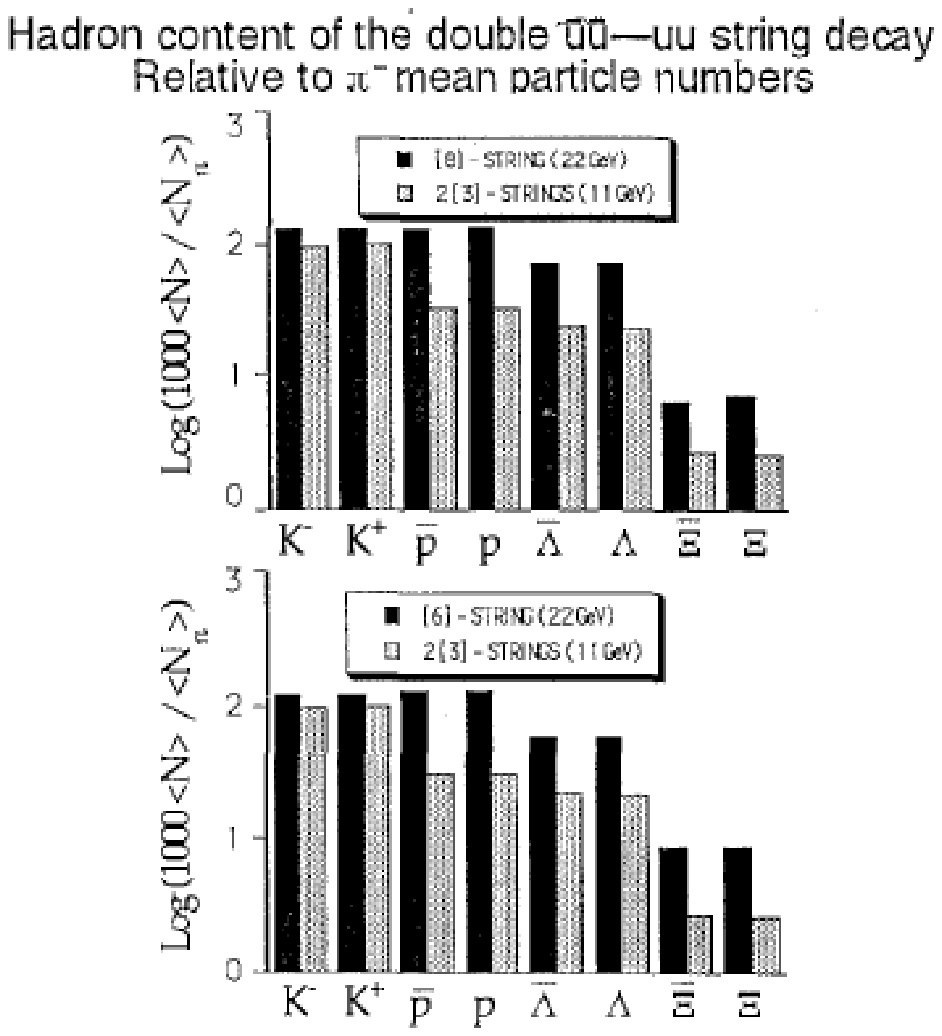}}
\vspace{0.2cm}
\caption{\label{fig1}Hadronic content of the decay of octet and sextet strings relative to the number of $\pi^{-}$.} 
\end{figure}

\begin{figure}
\resizebox{0.5\textwidth}{0.3\textheight}{
\includegraphics{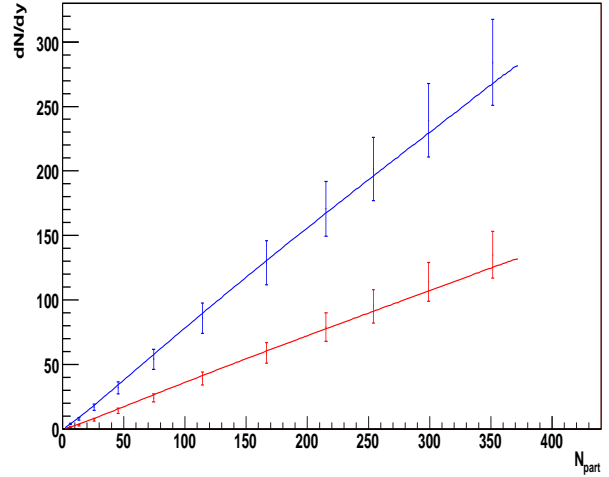}}
\vspace{0.2cm}
\caption{\label{fig2}$p_{T}$ integrated antiproton(red, X 10) and neutral pion (blue) spectra as a function of
  centrality compared to PHENIX data. }

\end{figure}

\begin{figure}
\resizebox{0.5\textwidth}{0.3\textheight}{
\includegraphics{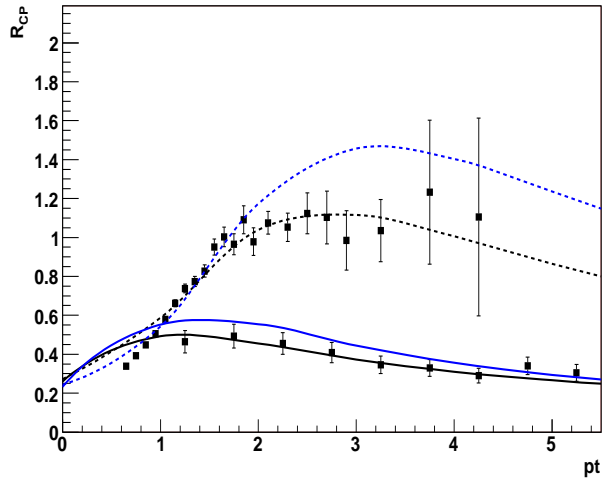}}
\vspace{0.2cm}
\caption{\label{fig3}$R_{CP}$(0-10\% central/60-92\% peripheral) for pions
  (solid line) and $(p+\overline{p})/2$ (dashed) compared to PHENIX data. In blue, LHC predictions. }

\end{figure}

\begin{figure}
\resizebox{0.5\textwidth}{0.3\textheight}{
\includegraphics{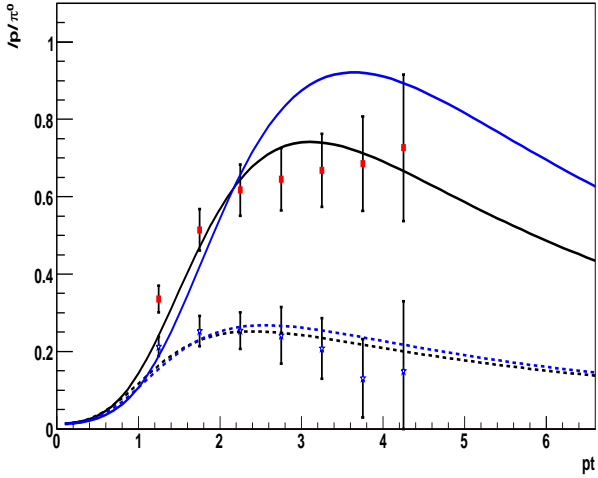}}
\vspace{0.2cm}
\caption{\label{fig4}Antiproton to neutral pion  ratio as a function of $p_{T}$ for for
  0-10\% (solid) and 60-92\% (dashed) centrality bins compared to PHENIX data. LHC predictions in blue. }

\end{figure}
\begin{figure}
\resizebox{0.5\textwidth}{0.3\textheight}{
\includegraphics{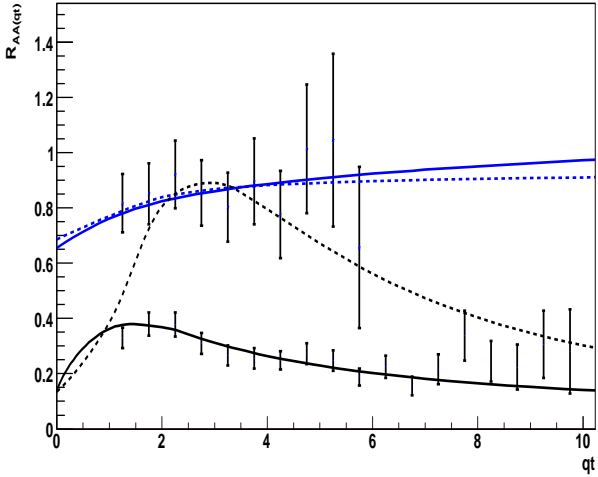}}
\vspace{0.2cm}
\caption{\label{fig5}Nuclear modification factor for neutral pions (solid) and
  protons (dashed) for 0-10\% central and 80-92\% peripheral bins compared
  to  PHENIX data. }

\end{figure}

\begin{figure}
\resizebox{0.5\textwidth}{0.3\textheight}{
\includegraphics{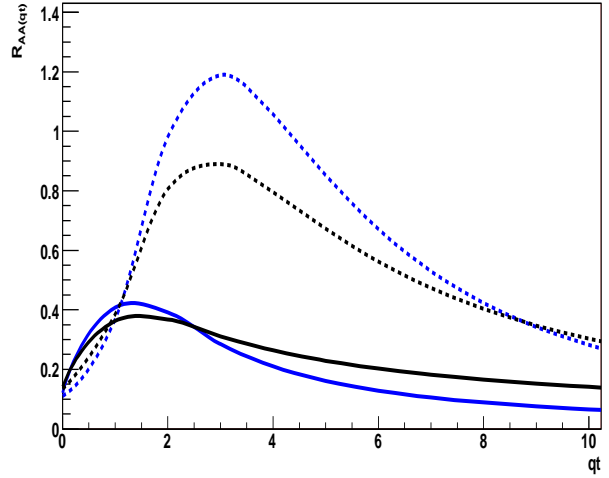}}
\vspace{0.2cm}
\caption{\label{fig6} Nuclear modification factor for 0-10\% central
  pions (solid) and protons (dashed) at RHIC (black) and LHC (blue)}

\end{figure}

\begin{figure}
\resizebox{0.5\textwidth}{0.3\textheight}{
\includegraphics{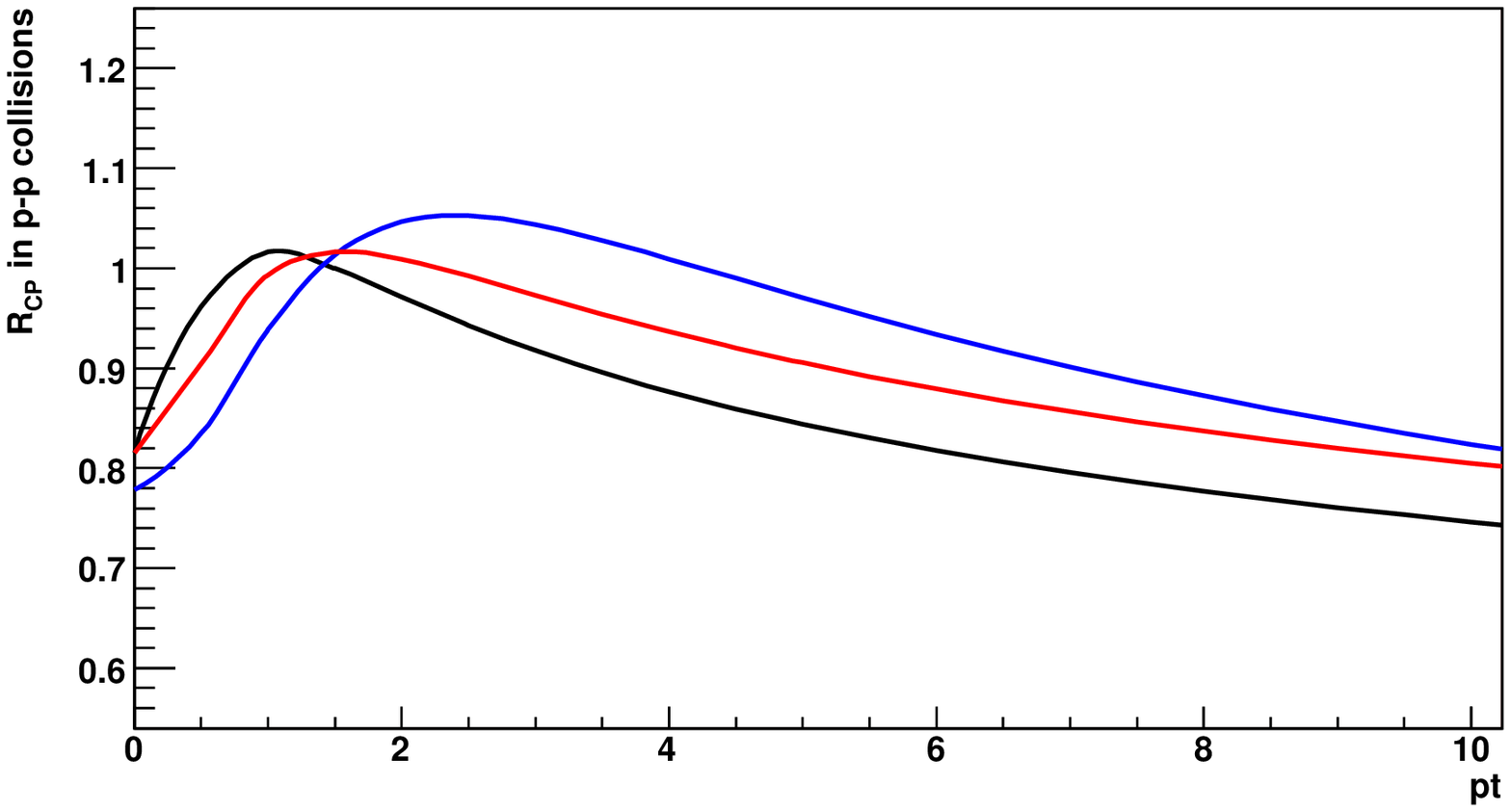}}
\vspace{0.2cm}
\caption{\label{fig7}Central to peripheral ratio for pp collisions at
  LHC. Black: $\pi^{0}$, Red: kaons, Blue: $\overline{p}$}

\end{figure}

%%%%%%%%%%%%%%%%%%%%%%%%%%%%%%%%%%%%%%%%%%%%%%%%%%%%%%%%%%%%%%%%%%%


\begin{thebibliography}{15}

 \bibitem{Armesto:1996kt}
  N.~Armesto, M.~A.~Braun, E.~G.~Ferreiro and C.~Pajares,
  %``Percolation approach to quark-gluon plasma and J/psi suppression,''
  Phys.\ Rev.\ Lett.\  {\bf 77} (1996) 3736;
  %%CITATION = PRLTA,77,3736;%%
  M.~Nardi and H.~Satz,
  %``String clustering and J/psi suppression in nuclear collisions,''
  Phys.\ Lett.\  B {\bf 442} (1998) 14.
  %%CITATION = PHLTA,B442,14;%%

 \bibitem{Braun:2000hd}
  M.~A.~Braun and C.~Pajares,
  %``Transverse momentum distributions and their forward-backward  correlations
  %in the percolating colour string approach,''
  Phys.\ Rev.\ Lett.\  {\bf 85} (2000) 4864;
  %%CITATION = PRLTA,85,4864;%%
  Eur.Phys.J C{\bf 85},349 (2000).

\bibitem{Dias de Deus:2003ei}
  J.~Dias de Deus, E.~G.~Ferreiro, C.~Pajares and R.~Ugoccioni,
  %``Universality of the transverse momentum distributions in the framework  of
  %percolation of strings,''
  Eur.\ Phys.\ J.\  C {\bf 40} (2005) 229.
  %%CITATION = EPHJA,C40,229;%%


\bibitem{Pajares:2005kk}
  C.~Pajares,
  %``String and parton percolation,''
  Eur.\ Phys.\ J.\  C {\bf 43} (2005) 9; J.~Dias de Deus and R.~Ugoccioni,
  %``Large P(T) Distributions At Rhic And Percolation Of Strings,''
  Eur.\ Phys.\ J.\  C {\bf 43} (2005) 249.
  %%CITATION = EPHJA,C43,249;%%

\bibitem{Adcox:2004mh}
  K.~Adcox {\it et al.}  [PHENIX Collaboration],
  %``Formation of dense partonic matter in relativistic nucleus nucleus
  %collisions at RHIC: Experimental evaluation by the PHENIX  collaboration,''
  Nucl.\ Phys.\  A {\bf 757} (2005) 184; E.~G.~Ferreiro, F.~del Moral and C.~Pajares,
  %``Transverse momentum fluctuations and percolation of strings,''
  Phys.\ Rev.\  C {\bf 69} (2004) 034901; J.~Dias de Deus and A.~Rodrigues,
  %``Transverse momentum fluctuations from clustering and percolation of
  %strings,''
  arXiv:hep-ph/0308011.
  %%CITATION = HEP-PH/0308011;%%

\bibitem{Alt:2006jr}
  C.~Alt {\it et al.}  [NA49 Collaboration],
  %``Centrality and system size dependence of multiplicity fluctuations in
  %nuclear collisions at 158-A-GeV,''
  Phys.\ Rev.\  C {\bf 75} (2007) 064904; L.~Cunqueiro, E.~G.~Ferreiro, F.~del Moral and C.~Pajares,
  %``Multiplicity fluctuations in the string clustering approach,''
  Phys.\ Rev.\  C {\bf 72} (2005) 024907.
  %%CITATION = PHRVA,C72,024907;%%

\bibitem{Amelin:1994mc}
  N.~S.~Amelin, M.~A.~Braun and C.~Pajares,
  %``String Fusion And Particle Production At High-Energies: Monte Carlo String
  %Fusion Model,''
  Z.\ Phys.\  C {\bf 63} (1994) 507.
  %%CITATION = ZEPYA,C63,507;%% 

 \bibitem{Armesto:2001et}
  N.~Armesto, C.~Pajares and D.~Sousa,
  %``Analysis of the first RHIC results in the string fusion model,''
  Phys.\ Lett.\  B {\bf 527} (2002) 92.
  %%CITATION = PHLTA,B527,92;%%
 
\bibitem{Mohring:1992wm}
  H.~J.~Mohring, J.~Ranft, C.~Merino and C.~Pajares,
  %``String Fusion In The Dual Parton Model And The Production Of Anti-Hyperons
  %In Heavy Ion Collisions,''
  Phys.\ Rev.\  D {\bf 47} (1993) 4142;
  %%CITATION = PHRVA,D47,4142;%%
  C.Greiner AIP Conf. Proc.644,337 (2003) nucl-th/0208080.

\bibitem{Hwa:2003ce}
  R.~C.~Hwa and C.~B.~Yang,
  %``Fractional energy loss and centrality scaling,''
  Phys.\ Rev.\  C {\bf 69} (2004) 034902;
  %%CITATION = PHRVA,C69,034902;%%
  V.~Greco, C.~M.~Ko and P.~Levai,
  %``Parton coalescence and antiproton/pion anomaly at RHIC,''
  Phys.\ Rev.\ Lett.\  {\bf 90} (2003) 202302;
  %%CITATION = PRLTA,90,202302;%%
  R.~J.~Fries, B.~Muller, C.~Nonaka and S.~A.~Bass,
  %``Hadronization in heavy ion collisions: Recombination and fragmentation  of
  %partons,''
  Phys.\ Rev.\ Lett.\  {\bf 90} (2003) 202303; 
  R.~J.~Fries, B.~Muller, C.~Nonaka and S.~A.~Bass,
  %``Hadron production in heavy ion collisions: Fragmentation and  recombination  %from a dense parton phase,''
  % R.~J.~Fries, B.~Muller, C.~Nonaka and S.~A.~Bass,
  %``Hadron production in heavy ion collisions: Fragmentation and  recombination  %from a dense parton phase,''
  Phys.\ Rev.\  C {\bf 68} (2003) 044902;
  L.~Maiani, A.~D.~Polosa, V.~Riquer and C.~A.~Salgado,
  %``Counting valence quarks at RHIC and LHC,''
  Phys.\ Lett.\  B {\bf 645} (2007) 138.
  %%CITATION = PHLTA,B645,138;%%

\bibitem{Adler:2003cb}
  S.~S.~Adler {\it et al.}  [PHENIX Collaboration],
  %``Identified charged particle spectra and yields in Au + Au collisions at
  %s(NN)**(1/2) = 200-GeV,''
  Phys.\ Rev.\  C {\bf 69} (2004) 034909.
  %%CITATION = PHRVA,C69,034909;%%



 \bibitem{Brodsky:1973kr}
  S.~J.~Brodsky and G.~R.~Farrar,
  %``Scaling Laws At Large Transverse Momentum,''
  Phys.\ Rev.\ Lett.\  {\bf 31} (1973) 1153;
  %%CITATION = PRLTA,31,1153;%%
  V.A.Matreev,R.M.Murddyan, A.N.Tavkheldize Lett. Nuuovo Cimento {\bf 7},
  719(1973); G.~P.~Lepage and S.~J.~Brodsky,
  %``Exclusive Processes In Perturbative Quantum Chromodynamics,''
  Phys.\ Rev.\  D {\bf 22} (1980) 2157.


\bibitem{Adler:2003qi}
  S.~S.~Adler {\it et al.}  [PHENIX Collaboration],
  %``Suppressed pi0 production at large transverse momentum in central Au +  Au
  %collisions at s(NN)**(1/2) = 200-GeV,''
  Phys.\ Rev.\ Lett.\  {\bf 91} (2003) 072301.
  %%CITATION = PRLTA,91,072301;%%

\bibitem{Albacete:2003iq}
  J.~L.~Albacete, N.~Armesto, A.~Kovner, C.~A.~Salgado and U.~A.~Wiedemann,
  %``Energy dependence of the Cronin effect from non-linear QCD evolution,''
  Phys.\ Rev.\ Lett.\  {\bf 92} (2004) 082001.
  %%CITATION = PRLTA,92,082001;%%

\bibitem{Topor Pop:2007hb}
  V.~Topor Pop, M.~Gyulassy, J.~Barrette, C.~Gale, S.~Jeon and R.~Bellwied,
  %``Transient field fluctuations effects in d+Au and Au+Au collisions at
  %sNN=200 GeV,''
  Phys.\ Rev.\  C {\bf 75} (2007) 014904;
  %%CITATION = PHRVA,C75,014904;%%
%  E.L.Bratkovskaya et al Phys.Rev.C{\bf 69}, 054904 (2004).
  E.~L.~Bratkovskaya {\it et al.},
  %``Strangeness dynamics and transverse pressure in relativistic nucleus
  %nucleus collisions,''
  Phys.\ Rev.\  C {\bf 69} (2004) 054907;
  %%CITATION = PHRVA,C69,054907;%%


\end{thebibliography}
\end{document}